# *Eco-Route*: Recommending Economical Driving Routes For Plug-in Hybrid Electric Vehicles

Yan Ding* , Yuanjian Zhang †
* College of Computer Science, Chongqing University, Chongqing 400044, China
† State Key Laboratory of Automotive Dynamic Simulation and Control, Jilin University, Changchun, 130022, China

*Abstract*—High fuel consumption cost results in drivers' economic burden. Plug-In Hybrid Electric Vehicles (PHEVs) consume two fuel sources (i.e., gasoline and electricity energy sources) with floating prices. To reduce drivers' total fuel cost, recommending economical routes to them becomes one of the effective methods. In this paper, we present a novel economical path-planning framework called *Eco-Route*, which consists of two phases. In the first phase, we build a driving route cost model (DRCM) for each PHEV (and driver) under the energy management strategy, based on driving condition and vehicles' parameters. In the second phase, with the real-time traffic information collected via the mobile crowdsensing manner, we are able to estimate and compare the driving cost among the shortest and the fastest routes for a given PHEV, and then recommend the driver with the more economical one. We evaluate the two-phase framework using 8 different PHEVs simulated in Matlab/Simulink, and the real-world datasets consisting of the road network, POI and GPS trajectory data generated by 559 taxis in seven days in Beijing, China. Experimental results demonstrate that the proposed model achieves good accuracy, with a mean cost error of less 8% when paths length is longer than 5 km. Moreover, users could save about 9% driving cost on average if driving along suggested routes in our case studies.

## I. INTRODUCTION

To alleviate pressure that the shortage of natural resources to the society, people from different territory have made a lot of effort. In the automotive industry, some methods have been proposed to reduce the dependent on fossil fuel. Among these raised methods, route planning is viewed as a reasonable method that can improve vehicle fuel economy, which is proved by previous works in traditional vehicles (normally is gasoline-burning vehicle) [1] [2]. The plug-in hybrid electric vehicle (PHEV) is another solution to improve the fuel economy. Despite the good performance of PHEV in vehicle fuel saving, the route planning for PHEV is still necessary to save fuel further [3] [4]. Therefore, this paper aims to come up with a method to plan routes, which can improve fuel economy for PHEV. PHEV, as a novel energy vehicle, is driven by two power resources: the internal combustion engine (ICE, normally is gasoline engine) and the rechargeable energy storage system (RESS, normally is battery). Compared with the traditional vehicle, different power sources in PHEV lead to complex energy management strategy, making the route planning for PHEV can be quite various [5].

In general, a PHEV can operate in the CD (charging depleting) or CS (charging sustain) stage, governed by energy management strategy according to different battery SOC [6] [7]. In CD stage, the battery is the primary energy source, and the engine would start to output energy when there is large tractive power or high speed requirement. In CS stage, on the contrary, engine plays the role of principal energy source. The battery outputs or recycles energy when there is the requirement to adjust the engine operation points and vehicle is driving at low speed. The switching from the CD stage to CS stage may result in different usage percentages of the two kinds of energy. In the route planning process related to the fuel saving, it is of vital importance to calculate the total energy cost. Hence, the disparate energy usage percentage causes different total energy cost, which has the obvious influence on the route planning [5]. In addition, When PEHV is operating in CD or CS stage, the working timing and status of the two power sources are determined by the energy management strategy too. The energy management would pick up the most optimal engine and motor operating combination (normally is the power combination), that can save fuel in the largest degree in each time interval. Different power combination leads to different energy usage percentage, affecting the route planning result.

Except the energy management strategy, the method, which unifies the fossil fuel usage and electricity usage to calculate the total energy consumption, has large influence on the route planning for PHEV. As is known, the unit of fossil fuel consumption is $g$, $L/100km$ or $MPG$, while the unit of electricity usage is $kWh$ or $J$. Different units of the two kinds of energy increase the difficulty of total fuel consumption calculation. In previous study about PHEVs' energy management strategy, people tried to convert electricity usage to the equivalent fossil fuel consumption to solve the problem [8].

Based on the upper discussion, we proposed a two-phase framework to plan the route for PHEV with the target that improving the fuel economy further. Firstly, we model and estimate the total driving cost by considering the impact from energy management strategy and unifying the two kinds of energy usage by monetary unit. Secondly, we calculate the driving cost of potential routes with a real-time traffic information via the mobile crowdsensing manner (MSC), and the recommend the most economical one for drivers. Thanks to ubiquitous availability of GPS (Global Positioning System) sensors in vehicles, GPS trajectory data that records the real driving conditions (e.g., speed and acceleration) can be easily gathered at a large scale, which makes MSC possible.

In this paper, we study the routing problem of PHEVs that aims at reducing the driving cost for users, via modeling the driving cost. To be specific, we make the following main contributions:

- We prove that the most economical driving route may be inconsistent between PHEVs and gasoline-burning vehicles from the source to the destination.
- We develop a two-phase framework called Eco-Route to model driving cost, and then recommend economical driving route for PHEVs with consideration of PHEV energy management strategy feature and calculating two kinds of energy usage by unified monetary unit.
- We leverage simulation data of 8 different PHEVs, the open data extracted from OpenStreetMap and the historical GPS trajectory data to evaluate the framework extensively. Experimental results demonstrate the effectiveness of the proposed two-phase framework in terms of modelling accuracy and providing the fuel-efficient driving routes.

The rest of the paper is organized as follows. In Section II, a contrast study is presented to verify a view, the most economical driving route may be inconsistent between PHEVs and gasoline-burning vehicles. Then, we introduce some basic concepts, and present the details about the two-phase framework of our proposed *Eco-Route* system in Section III. Next, we evaluate the performance of the proposed framework in Section IV. In Section V, we review the related work, and show how this paper differs from previous research. Finally, we conclude the paper and propose the future directions in Section VI.

## II. A CONTRAST STUDY

In this Section, we present a contrast study between a gasoline-burning vehicle and a PHEV. The study provides readers with a proof of two concepts:

- Two types of vehicles probably have different fuel-efficient routes from the source to the destination.
- The value of SOC is highly correlated with the fuel cost of PHEVs.

To accomplish the contrast study, we have two steps. Firstly, we collect the GPS trajectory of the shortest route and fastest route from a source to the destination via MSC. Secondly, we calculate the energy cost of the PHEV and gasoline-burning vehicle when traveling on these two routes. For the gasoline-burning vehicle, we estimate the consumed gasoline cost via inputing the GPS trajectory of two routes into our prior models, called *GreenPlanner* [1]. For the PHEV, we calculate its fuel (i.e., gasoline and electricity energy) cost via inputing the GPS trajectory into the simulated vehicle model in Matlab/Simulink. We compare the fuel cost along the two driving routes for two cars, and the result is shown in Table.I and Fig.1.

Table.I presents the quantitative information (i.e., the amount of gasoline and the cost) of these two routes for the gasoline-burning vehicle. As shown in this table, the fastest driving route is more economical in this case. Fig.1 shows the driving cost of two routes for a PHEV. X-axis denotes different SOC values, and y-axis denotes the driving route cost. We can see that the most economical driving route for a PHEV is highly correlated with the value of SOC. The economical driving route for a PHEV is different from that for a gasoline-burning vehicle (e.g., when SOC is more than 0.6, the economical route is the shortest route, not the fastest route). From the contrast study, we have proved that *previous work of finding fuel-efficient routes for gasoline-burning cars can not be applied directly in PHEVs*. Thus, it is crucial to model the driving route cost of PHEVs, and suggest the economical routes for drivers. Moreover, we should take into account of the SOC value of PHEVs when modeling.

**Table I:** The driving route cost (i.e., *gasoline cost*) about the fastest and shortest driving routes for a *gasoline-burning vehicle* ($Toyota\ Camry$).

| Route Type | Gasoline ($L$) | Cost ($CNY$) |
|---|---|---|
| Fastest route (*Eco*) | 0.841 | 5.21 |
| Shortest route | 1.047 | 6.49 |

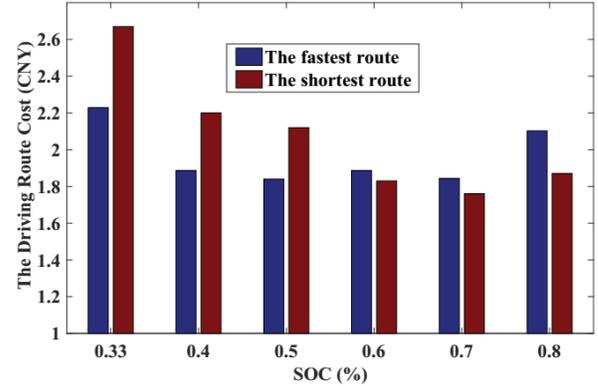

**Figure 1:** The driving route cost (i.e., *gasoline and electricity energy cost*) of the fastest and shortest driving routes for a *PHEV* ($Hyundai\ Sonata$) with varied battery SOC.

## III. THE TWO-PHASE FRAMEWORK

In this section, we will firstly introduce some basic concepts. Secondly, we will build the **d**riving **r**oute **c**ost **m**odel (**DRCM**) to estimate the potential fuel cost, when a PHEV has travelled a given path. Finally, we will estimate the fuel cost of shortest and fastest route, and recommend a more economical route for the driver.

### A. Basic Concepts

**Definition 1** (GPS Trajectory). *A GPS trajectory is a sequence of time-ordered spatial points $(p_1 \rightarrow p_2 \rightarrow \cdots \rightarrow p_9)$, where each point has a geospatial coordinate, the instantaneous speed(in km/h), the distance that has been covered (in km), and a timestamp, denoted by $p_i = (t_i, latitude_i, longitude_i, v_i)$.*

**Definition 2** (Equivalent Consumption Minimization Strategy). *Equivalent consumption minimization strategy (ECMS) is proposed for the power management of PHEV, which is* one of common energy management methods. The electricity consumption can be converted to an equivalent amount of gasoline by the equivalent factor ($s$) of ECMS, and instantaneous total power split ratio can be calculated by minimizing the equivalent fuel consumption. In the paper, $s$ of ECMS is 2.35.

**Definition 3** (Gasoline Consumption Reading). *The gasoline consumption reading of a PHEV is a data log, which records the information about the total gasoline usage (in gram) of the vehicle at the sampling time $t$, denoted by $m_f(t)$.*

**Definition 4** (Electricity Consumption Reading). *A electric meter reading of a PHEV at the sampling time $t$ is also a data log, recording the information about the total electricity usage (in kW-second) of the vehicle, denoted by $p_{batt}(t)$.*

*Note*: Both the gasoline and the electricity consumption reading of PHEVs in our experiment can be obtained by inputting GPS trajectory of drivers into simulated vehicle models in Matlab. The readings can be considered as the actual fuel consumption. We will evaluate the effectiveness of the simulated vehicle model in Section IV-B to make sure the accuracy of the readings.

*B. Phase I: Driving Route Cost Model (DRCM) Building*

The objective of DRCM is to estimate the potential amount of total fuel cost (i.e., *the cost of gasoline and electricity energy*), when a driver has travelled a certain path under the driving condition (e.g., *the vehicle speed and acceleration*) recorded by the GPS trajectory data.

The general driving route cost $drc(t)$ of a PHEV at any time $t$ (CD or CS stage) can be represented by Eqn.1 [9] [10]. The first part in the equation is gasoline cost, and the second part is the electricity cost.

$$drc(t) = n \times \underbrace{\frac{m_f(t)}{\rho}}_{gasoline\ cost} + m \times \underbrace{\frac{p_{batt}(t)}{3600}}_{eletricity\ cost} \quad (1)$$

where $n$ ($CNY/L$) and $m$ ($CNY/kWh$) is the current price of gasoline and electric energy respectively, $\rho$ denotes the fuel density, $p_{batt}(t)$ ($kWs$) denotes the instantaneous battery power, $m_f(t)$ ($g$) is the fuel mass flow rate, which can be calculated in Eqn.2.

$$m_f(t) = \frac{p_{eng}(t)}{\eta_{eng}(t) Q_{lhv}} \quad (2)$$

where $Q_{lhv}$ ($MJ/kg$) is the fuel lower heating value (energy content per unit of mass), $\eta_{eng}$ is the engine efficiency, $s$ is the equivalent factor, $p_{eng}(t)$ ($kWs$) is the engine power at time $t$.

Therefore, the the driving route cost $drc(t)$ in Eqn.1 can also be expressed as:

$$drc(t) = n \times \frac{p_{eng}(t)}{\eta_{eng}(t) Q_{lhv} \rho} + m \times \frac{p_{batt}(t)}{3600} \quad (3)$$

In the Eqn.3, only $p_{eng}(t)$ and $p_{batt}(t)$ is variant. To obtain the $drc(t)$ of a given path with the GPS trajectory under a PHEV's ECMS, we mainly take three steps to calculate $p_{eng}(t)$ and $p_{batt}(t)$. The first step is to calculate the total amount of $p_{eng}(t)$ and $p_{batt}(t)$ based on our prior model. The second step is to identify the numerical relationship between $p_{eng}(t)$ and $p_{batt}(t)$ based on ECMS. The final step is to calculate the driving route cost.

*1) Step 1, Calculating the total amount of $p_{eng}(t)$ and $p_{batt}(t)$*: $p_{eng}(t) + p_{batt}(t)$ can be calculated by Eqn.4.

$$p_{eng}(t) + p_{batt}(t) = f(t) \times v(t) \quad (4)$$

where $f(t)$ and $v(t)$ is the total driving force and vehicle speed, respectively.

Fuel is consumed to generate the driving force $f(t)$ to overcome the total resistance, including the friction resistance caused by the road $f_f(t)$, the gradient resistance $f_g(t)$ caused by the gravity, the air resistance $f_r(t)$ and the acceleration resistance $f_a(t)$. The relationship of the above forces can be represented in Eqn.5.

$$f(t) = f_f(t) + f_g(t) + f_r(t) + f_a(t) \quad (5)$$

The road frictional force $f_f(t)$ can be characterized by the gravitational force acting on the vehicle and estimated according to Eqn.6.

$$f_f(t) = \mu m g \cos(\theta) \quad (6)$$

where $\mu$ is the coefficient of friction, $m$ is the mass of the vehicle, $g$ is the gravitational acceleration, and $\theta$ is the ground slope of the road.

The gravitational force $f_g(t)$ can be approximated using the following equation:

$$f_g(t) = m g \sin(\theta) \quad (7)$$

The force caused by the air resistance $f_r(t)$ can be estimated by the following equation:

$$f_r(t) = \frac{1}{2} \varphi C_r A v^2(t) \quad (8)$$

where $\varphi$ is the coefficient of the air resistance, $A$ is the frontal area of the car, $C_r$ is the air density, and $v$ is the speed of the vehicle at time $t$.

The force caused by the acceleration force $f_a(t)$ can be calculated by Eqn.9 [1].

$$f_a(t) = \beta m a(t)(\gamma + |path.ts| + |path.n| + |path.poi|) \quad (9)$$

where $a(t)$ and $v(t)$ is the acceleration and the speed of the vehicle at the time $t$, respectively; $\gamma$ is a constant coefficient which is used to characterize the case of traffic congestions; $|path.ts|, |path.n|, |path.poi|$ correspond to the number of traffic lights/stop signs, the number of neigbouring edges, and the number of specific POIs of the given path respectively. Thus, the driving force $f(t)$ can be expressed by the Eqn.10.

$$f(t) = \mu m g \cos(\theta) + m g \sin(\theta) + \frac{1}{2} \varphi C_r A v^2(t)$$
$$+ m a(t)(\gamma + |path.ts| + |path.n| + |path.poi|) v(t) \quad (10)$$

The total fuel consumption rate can be expressed in Eqn.11 based on Eqn.4 and 10.

$$p_{eng}(t) + p_{batt}(t) = k_1 v(t) + k_2 v(t)^3 \\ + (k_3 + k_4|path.ts| + k_5|path.n| + k_6|path.poi|)v(t)\Delta v \quad (11)$$

where $\Delta v$ is speed difference at the sampling time interval of GPS trajectory data; $k_1$, $k_2$, $k_3$, $k_4$, $k_5$ and $k_6$ are constant coefficients but vary among different PHEVs, which can be easily derived from the GPS data in Matlab/Simulink. We can easily calculate the total amount of $p_{eng}(t) + p_{batt}(t)$ at any time based on the derived coefficients. Thus, $p_{eng}(t) + p_{batt}(t)$ can also be represented in Eqn.12. The information about the number of traffic lights/stop signs, the neighbouring edges, and the specific POIs of the path can be extracted from OpenStreetMap (OSM) platform.

$$p_{eng}(t) + p_{batt}(t) = Q \quad (12)$$

*2) Step 2, Conducting Equivalent Consumption Minimization Strategy (ECMS):* This step is mainly to investigate the relationship between $p_{eng}(t)$ and $p_{batt}(t)$ based on ECMS. To accomplish the ideal fuel consumption, PHEV conducts a reasonable energy management strategy to spilt the power requirement $Q$ between the engine and the battery. For example, $p_{eng}(t) = \mu * Q$, $p_{batt}(t) = (1 - \mu) * Q$. $\mu$ is the power split ratio, determined by ECMS. Though the strategy aims to minimize the energy consumption, it ignores the current price of two energy. When recommending the economical driving route for driver, we should take into account of the current energy price. Among the numerous proposed energy management solutions, the Equivalent Consumption Minimization Strategy (ECMS) is an instantaneous method which could offer well performance. ECMS aims to find the best power spilt ratio between the engine and battery to achieve minimum equivalent fuel consumption at time $t$, which can be normally expressed in Eqn.13 [11].

$$\arg\min[m_f(t) + m_e(t)] \quad (13)$$

Where $m_f(t)$ is the fuel mass flow rate, $m_e(t)$ means the additional fuel consumption transformed from the electric energy consumption. The $m_e(t)$ can be calculated in Eqn.14:

$$m_e(t) = \frac{s}{Q^{lhv}} p_{batt}(t) \quad (14)$$

where $s$ is a fixed equivalent factor. The equivalent factor is the core of the ECMS and normally varies from 2.4 to 2.7. In this paper, it's value is 2.5. Therefore, based on Eqn.2 and 14, Eqn.13 can be rewritten as:

$$\arg\min[\frac{p_{eng}(t)}{\eta_{eng}(t)Q_{lhv}} + \frac{s}{Q_{lhv}} p_{batt}(t)] \quad (15)$$

*3) Step 3, Calculating Driving Route Cost:* In step 1, we calculate the total amount of $p_{eng}(t)$ and $p_{batt}(t)$. In step 2, we measure the numerical relationship between $p_{eng}(t)$ and $p_{batt}(t)$. In this step, based on step1 and 2, we can calculate the value of $p_{eng}(t)$ and $p_{batt}(t)$, respectively in Eqn.16 and Eqn.17.

$$p_{eng}(t) + p_{batt}(t) = Q \\ \arg\min[\frac{p_{eng}(t)}{\eta_{eng}(t)Q_{lhv}} + \frac{s}{Q_{lhv}} p_{batt}(t)] \quad (16)$$

$$\Downarrow \\ \begin{cases} p_{eng}(t) = N \\ p_{batt}(t) = M \end{cases} \quad (17)$$

Thus, we can calculate the $drc(t)$ in Eqn.18 based on Eqn.1 and 17.

$$drc(t) = n \times \frac{N}{\eta_{eng}(t)Q_{lhv}\rho} + m \times \frac{M}{3600} \quad (18)$$

Therefore, the total driving route cost $DRC$ of a given path for a PHEV can be modeled by integrating the total driving cost rate $drc(t)$ during the time period $[t_1, t_n]$, as follows.

$$\hat{D}RC = \int_{t_1}^{t_n} \hat{drc}(t) dt \quad (19)$$

where $t_1$ and $t_n$ refer to the time when the PHEV enters and leaves the given path, respectively.

Compared to the ground truth $drc(t)$ that is obtained by simulation, each piece of $\hat{drc}(t)$ may either be overestimated or underestimated, thus the modelling error could be accumulated or reduced as the path becomes longer and contains more GPS points. We will investigate the modelling performance of DRCM on paths under different path lengths quantitatively in Section IV-C.

*C. Phase II: Economical Driving Route Recommending*

With the model constructed, we are able to estimate the cost of a driving route, given his/her GPS trajectory data. However, the model can only operate ex-postly (i.e., compute the energy cost after the trip is completed). In another word, because of the unavailability of the GPS trajectory data on each potential edge from the source to the destination, recommendation of economical routes cannot be provided to drivers. To migrate the problem, we propose a three-step procedure, the technical details of which are discussed as follows.

*1) Real-time Traffic Information Collection:* The core of the real-time traffic information collection is to predict the information about GPS trajectory (e.g., the number of sampling points, the speed information of each point) on an edge if the driver would travel it, which is proved to be quite challenging. Here, to simplify the issue, *we assume that we just have abundant real-time GPS trajectory data generated by other drivers on each edge during a time period $T$*[1]. The data can be input into the **DRCM** and we will get the potential cost on that edge for the given PHEV. We believe that the assumption is reasonable, as in recent years there are an increasing number of mobile crowdsensing apps that share real-time GPS trajectory data and traffic information, such as Waze[2].

---

[1] In this paper, $T$ is set to 1 hour and 2 hours during the rush time hours and non-rush time hours, respectively

More specifically, for an edge, after the driver submits his request at time $t_0$, we first retrieve all the real-time GPS trajectory data contributed by all passing-by drivers on that edge during the time period $[t_0, t_0 - T]$. We then keep the GPS trajectory data with the most similar cumulative travel distance ($\sum_{i=1}^{n} \Delta d_i$, where $n$ is the number of GPS points on the edge) to the actual edge length and use it to represent the case if the given driver travells on it. The travel time on the edge can also be easily obtained. The selected GPS trajectory data is denoted as a pair of *driverId* and *edgeId*, e.g., $\langle dr_i, e_j \rangle$. The effectiveness of collecting real-time information via the Mobile Crowdsensing (MCS) manner for route-recommending will be evaluated in Section IV-D1. *Note, for our own simulation, the sampling time of the vector sequences is 1 s. However, the sampling time of GPS trajectory data providing by actual vehicles may much longer (i.e., 6 s). We approximate the missing data by linear interpolationp.*

*2) Driving Route Cost Estimation:* With the real-time traffic information collected, based on the built **DRCM** for each PHEV, we are able to estimate the potential cost on a path ($Path_i$) to be travelled for the given PHEV ($phev_g$) by summing the cost on all edges ($e_k|_{k=1}^{N}$) belonging to by the path, as shown in Eqn. 20.

$$\hat{DRC}(phev_g, path_i) = \sum_{k=1}^{N} \hat{drc}(phev_g, \langle phev_o, e_k \rangle) \quad (20)$$

*3) Economical Driving Route Recommending:* From a source (S) to a destination (D), to obtain the most economical route for a given PHEV, an intuitive idea is to enumerate all possible paths, then estimate and compare the potential cost for each path based on the two steps discussed earlier (i.e., Real-time Traffic Information Collection and Driving Route Cost Estimation) for each path, and pick up the one with the least fuel as the recommendation to the driver. However, the intuitive solution is not practical because discovering all possible paths for a given source-destination (SD) pair is a well-know NP-hard problem. Considering the fuel consumption is highly correlated with the length of the route, and the vehicle's speed, we only compare the cost of the shortest and fastest routes from $S$ to $D$, recommending a more fuel-optimal route [12] [13].

## IV. EVALUATION

### A. Experimental Setup

*1) Data Preparation:* In our experiments, the gasoline and electricity consumption reading is generated by inputing GPS trajectory, providing by 595 taxis in seven days (2016) in Beijing city, into Matlab/Simulink. We use 30% total GPS trajectory data to train the **DRCM** of each PHEV, and use the rest data to evaluate the model accuracy. Some basic information about used PHEVs can be found in Table.II, including the car number, make, model, battery capacity and vehicle's mass. *Note that some necessary data pre-processing techniques such as denoising, data correction/interpolation, map matching on the GPS trajectory are applied before the modeling.*

[2]https://www.waze.com/

**Table II:** Main parameters of PHEVs used in the experiment.

| No. | Car Make | Car Model | Battery Capacity[1] | Mass[2] |
|---|---|---|---|---|
| 1 | Hyundai | Sonata | 7.6 | 1650 |
| 2 | Audi | A3 | 9.0 | 1620 |
| 3 | Benz | C350el | 6.3 | 1925 |
| 4 | BYD | Qin | 15.0 | 1677 |
| 5 | Saic | Rowe e950 | 12.0 | 1835 |
| 6 | VW | Golf | 9.0 | 1700 |
| 7 | Volvo | Volvo | 11.1 | 1996 |
| 8 | Saic | Rowe 550 | 12.0 | 1699 |

[1] Measured in $kWh$.
[2] Measured in $kg$.

*2) Evaluation Metric:* One metric, i.e., the mean error, is defined to evaluate the modeling accuracy.

$$error_i = \frac{\hat{DRC}(path_i) - DRC(path_i)}{DRC(path_i)} \times 100 \quad (21)$$

$$m\_error = \frac{\sum_{i=1}^{m} error_i}{m} \quad (22)$$

where $\hat{DRC}(path_i)$ is the fuel cost on the given path estimated via the built model; $DRC(path_i)$ is the actual fuel cost on that path, which is obtained by simulation.

### B. Evaluation on PHEV Simulation

In this paper, we take advantage of PHEV models in Matlab/Simulink under the given real speed sequences. We can obtain the fuel consumption data (i.e., gasoline and electricity energy) of PHEVs by simulation result, instead of collecting the data in real driving environment. To guarantee the accuracy of evaluation on **DRCM**, we should firstly verify the effectiveness of the simulated PHEV models based on engineering intuition. The method is that we compare the absolute velocity difference between the vehicle velocity sequence acquired in simulation and real driving condition. If the speed difference is less, then it is more safe to claim that the PHEV models is effective. Otherwise, it is ineffective. The result is shown in Table.III, and we can see that the accurate rate is 97.8% when the absolute velocity difference is less than 2 $m/s$. According to the experiment results, it is can be concluded that the model accuracy is reasonable and can be applied to evaluate the two-phase framework in this paper.

**Table III:** Quantitative information of the absolute velocity difference between the simulated cycle and the actual driving cycle.

| Absolute Velocity Difference (m/s) | The percentage of the total number |
|---|---|
| ≤ 1 | 92.7% |
| ≤ 2 | 97.8% |
| ≤ 3 | 99.3% |
| ≤ 4 | 99.9% |

### C. Evaluation on DRCM

We have ensured the accuracy of simulated PHEVs model in the above section. In this section, we want to know the performance of **DRCM**. In order to evaluate the model accuracy and how models perform on paths with different lengths qualitatively at different battery SOC, we bin the paths based on their length and compute the m_error of a certain PHEV. The corresponding results are shown in Fig.2. X-axis denotes the path length, and y-axis denotes the mean accuarcy. From the figure, we can see that, when the initial battery SOC is 0.8 or 0.33, the model accuracy gets higher as path length increases. In more detail, DRCM achieves a mean driving route cost error around 7% when the path length is 10 km. Moreover, **DRCM** achieves a quite stable mean error of less than 5% when the path length is longer than 15 km. Compared with the initial battery SOC is 0.8 or 0.5, the mean error with initial battery SOC = 0.33 is larger in most cases. This is because vehicle in CS stage (SOC = 0.33) would be mainly driven by engine. The engine, however, is the component with large inertia, so it may not fulfill the required power output as soon as possible, increasing the fuel consumption and resulting in larger mean error. Another obvious phenomenon is that the mean error with initial battery SOC = 0.5 dose not decrease with the path length increasing. As explained, the vehicle with initial battery SOC = 0.5 would encounter the mode switching from CD to CS, which means the transformation of the principle power sources: from battery to engine. As above is discussed, engine's larger inertia may cause larger fuel consumption and mean error. Generally, **DRCM** achieves a good performance based on the experiment results.

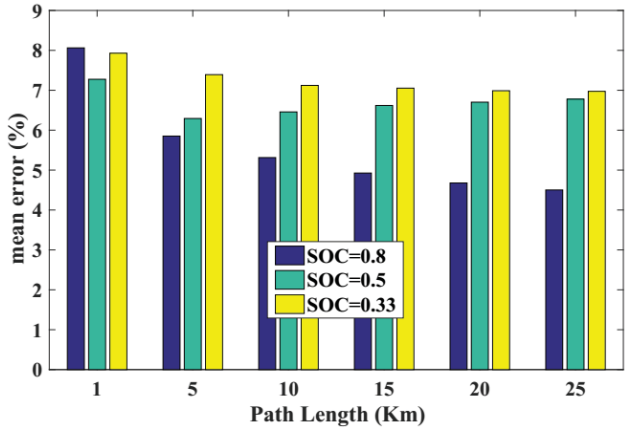

**Figure 2:** Model performance of **DRCM** by varing the length of the path when the PHEV model is *Sonata*.

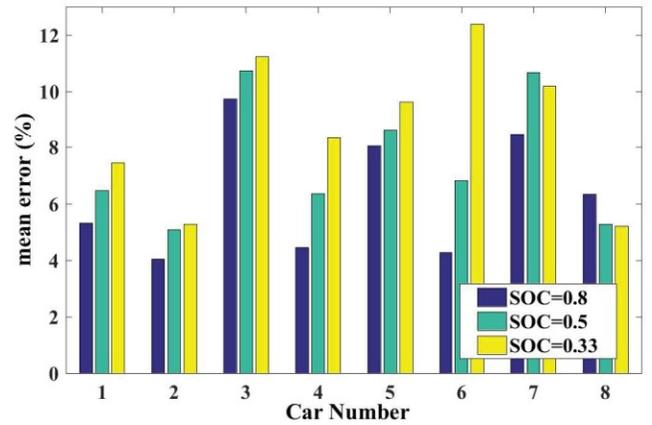

**Figure 3:** Model performance of different PHEVs at different battery SOC values when the path length is 10km.

We use the certain PHEV to evaluate **DRCM**, and also want to know the mean error of other PHEVs, when the path length is 10 km, and SOC is 0.8, 0.5 or 0.33.

The evaluation results of all PHEVs are shown in Fig.3. X-axis denotes the car number shown in Tabel.II, and y-axis denotes the mean error of PHEVs. Most mean errors of PHEVs are less than 10%. However, the mean error of No. 3, 5, 6 and 7 are higher than that of the other vehicles. This is because the vehicle mass of No. 3, 5, 6 and 7 are relatively larger than the others. Larger vehicle mass means bigger inertia, increasing the difficulty when simulated model follows the driving requirement [14].

### D. Evaluation on Path-Planning Algorithm

In the above section, we have ensured the accuracy of **DRCM**, and we will evaluate the path-planning algorithm sequentially. To evaluate the path-planning algorithm, we take two steps. The first step is to evaluate the effectiveness of our proposed method of collecting real-time traffic information. The second step is to evaluate the potential fuel consumption savings.

*1) Effectiveness of Real-time Traffic Information Collection:* For a given path with recorded GPS trajectory, we know the actual fuel consumed (obtained by simulation) on it, which can be used as the ground truth to compare with. Meanwhile, at the given starting time of the path, with the proposed collection method, we are also able to collect the the potential GPS trajectory data on each edge of the path. Then, based on PHEVs' **DRCM**, the amount of the potential fuel consumption can be estimated according to Eqn.20, when the vehicle travels from the source to the destination. If the estimation result is close to the ground truth, then it is safe to claim that the method of collecting real-time information is effective. Otherwise, it is ineffective.

We compute the m_error for all studied paths with length from 1 km to 25 km based on the real-time information collected via MCS. The results are presented in Fig.5. X-axis denotes the path length, and y-axis denotes the mean error. As a reference, results based on **DRCM** with the actual GPS trajectory data as input are also provided. We can see that with the real-time information collected via MCS, **DRCM** achieves an acceptable accuracy, with a mean absolute error less than 11% when the path length is 5 km, and an increasingly better accuracy when the path length increases. As expected, the estimation accuracy using the real-time information is worse than the corresponding one when the actual GPS trajectory data is used under all path lengths.

*2) Potential Savings On Driving Cost:* We are also interested in the potential savings on fuel, if drivers travel on our suggested routes. In this study, four real cases and a certain

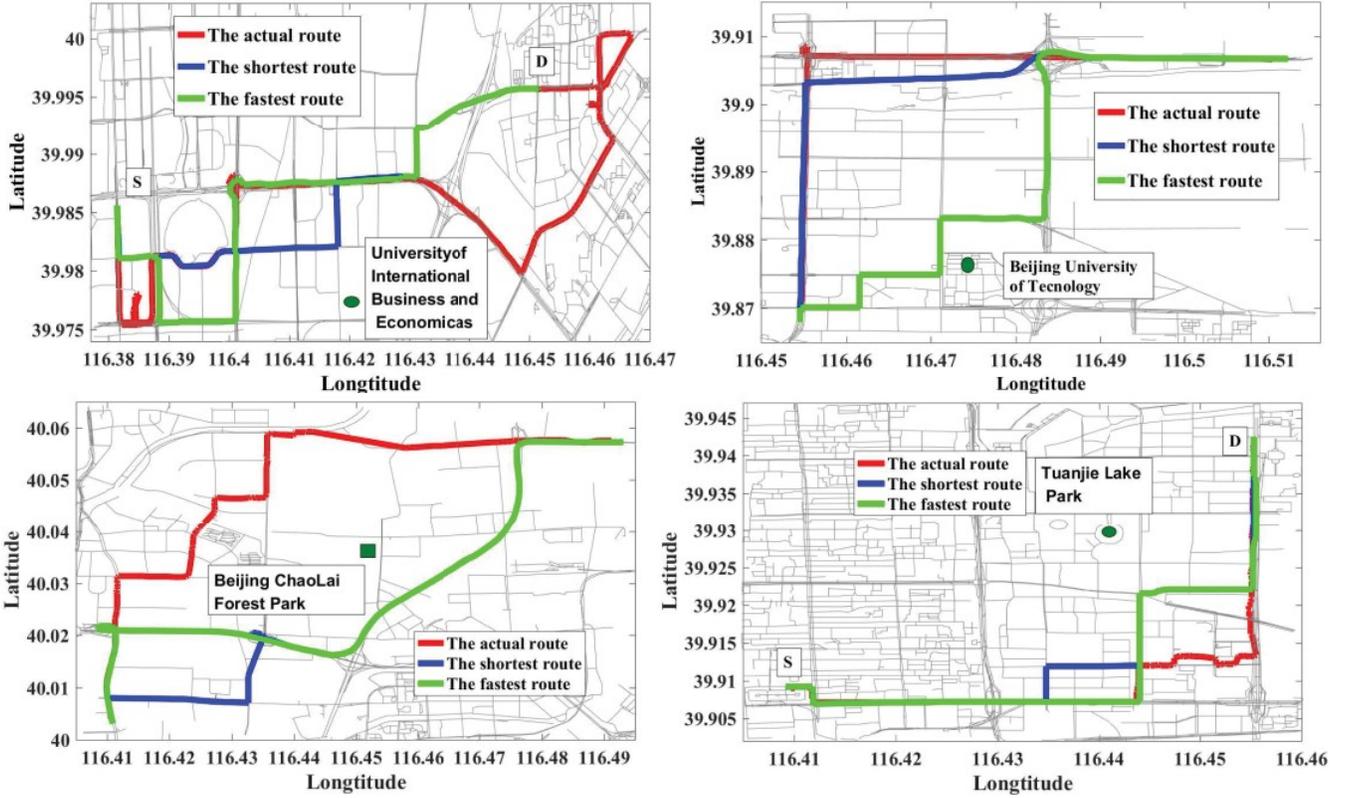

**Figure 4:** Comparison results between the actual, fastest and shortest driving routes for four cases, respectively (SOC = 0.8, *Sonata*). Case I (upper left); Case II (upper right); Case III (lower left); Case IV (lower right)

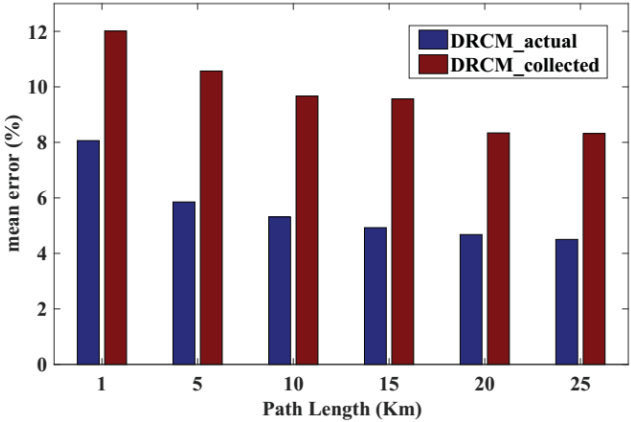

**Figure 5:** Results of the effectiveness of the real-time collection method under varied path lengths when the PHEV model is *Sonata* and SOC is 0.8.

PHEV model are selected when the SOC is 0.8. The actual, shortest and fastest driving routes in cases are shown in Fig.4. We further provide the quantitative information about these routes of the four cases in Table.IV. The potential savings in table on the fuel cost is defined as follow:

$$Saving = \frac{shortest.cost - fastest.cost}{min\{shortest.cost, fastest.cost\}} \quad (23)$$

From the table, we can see that the economical route for case I and III is the shortest route, whereas, for cases IV it is the fastest route. Hence, picking the shortest or fastest routes consistently is not fuel-optimal. In case II, the cost of the shortest and fastest route is equal, but these two routes are more fuel-optimal than the actual route. The average savings by choosing fuel-efficient routes in cases are around 9% on average as summarized in the table.

## V. RELATED WORK

In this section, we explained the related work below and point out how our work differs from the previous research work. In previous research, people have generally agreed that planning reasonable travel path can improve the fuel economy of the vehicle. To finding ideal route that can save fuel, researchers have done amount of work specific to the gasoline-burning vehicle. When people plan the economical route for driver, they may seek the fastest or shortest route as the most economical route. Ichimori et al. [15] tried to finding the shortest path for a vehicle when the vehicle has a limited capacity and is allowed to stop and refuel at certain locations. Sanders and Schultes [16] and Geisberger et al. [17] propose some hierarchy algorithms which run faster in real road network. These work, generally, seldom talk about the relationship between the fastest route and shortest route and its influence on fuel economy improvement. In this paper, we go a step further. We discussed different fuel consumption by fastest and shortest route, choosing the more economical one between them. In addition, although some researchers pay attention to plan the route for novel energy vehicle (i.e., PHEV), they seldom take the PHEV energy management feature into account. Artmeier et al. [18] propose an routing algorithm for EVs with the constraints of battery capacity and negative weight. Hausler et al. [19] provide a stochastic balancing algorithm to reduce the potential for excessively long queues at some charging stations. When we plan the route for PHEV, we consider the gasoline and electricity cost together, which is resulted from the energy management strategy, making the route planning more reasonable.

Table IV: Quantitative information about the suggested driving routes and the actual driving routes.

| Case | Actual Route | | Fastest Route | | Shortest Route | | Potential |
|---|---|---|---|---|---|---|---|
| | Distance(*km*) | Cost(*CNY*) | Distance(*km*) | Cost(*CNY*) | Distance(*km*) | Cost(*CNY*) | Savings(%) |
| I | 18.78 | 4.39 | 9.60 | 4.75 | 11.03 | 4.18 (*Eco*) | 4.8 |
| II | 10.70 | 2.92 | 10.39 | 2.49 (*Eco*) | 10.89 | 2.49 (*Eco*) | 14.7 |
| III | 14.75 | 3.40 | 14.29 | 3.40 | 14.70 | 2.99 (*Eco*) | 12.1 |
| IV | 9.54 | 2.46 | 9.24 | 2.34 (*Eco*) | 9.25 | 2.79 | 4.1 |

## VI. CONCLUSION

In this paper, we present a novel framework called *Eco-Route* to address the issues of the fuel consumption modelling and the fuel-efficient path-planning for PHEVs. In the modelling phase, we build DRCM for each PHEV based on the real driving condition along the path recorded by GPS trajectory data, actual fuel consumption (calculated in Matlab/Simulation). In the second phase, with the real-time traffic information collected via MSC, we estimate the fuel cost of shortest and fastest driving route based on DRCM of PHEVs, and then recommend a more fuel-efficient route for drivers. The proposed two-phase framework had been extensively evaluated using real-world datasets, which consists of road network, POI and the GPS trajectory data generated by 559 taxis in seven day in Beijing, China. Experimental results demonstrate that, our model achieves a good accuracy for different PHEVs, and users could save a considerable amount of fuel if driving along our suggested routes in four cases. In the future, we plan to broaden and deepen this work in several directions. Firstly, to better understand our proposed model, we plan to analyze the energy distribution under the ECMS. Secondly, we intend to build general DRCMs applied for any PHEVs.